# Defining Computer Art: Methods, Themes, and the Aesthetic Problematic

Tianhua Zhu (Shanghai Academy of Social Sciences)

**Abstract:** The application of computer technology in the field of art has given rise to novel modes of artistic practice, including media art, and it is a necessity to find a commensurable conceptual fundament. Therefore, computer art starting from the 1950s reenters the view. To clarify the definition, major methods for defining are reviewed, and it is argued that the thematic definition guided by situational logic provides a feasible approach. There is a triad of themes: the relationship between art and technology, the problem of machine creation, and the ontology of art. Consisted of primitive and mutually supportive questions, a logical space of questioning, i.e. a problematic, is formed as the basis for the identity of computer art. Among them, the problem of the ontology of art is located at the logical starting point of questioning, and this problematic is therefore an aesthetic one. The anticipation that computer art presents and responds to the above-mentioned aesthetic problematic suggests the plausibility of computer art being a legitimate category of art.
**Bio:** Zhu Tianhua, Ph.D. in Aesthetics, assistant research faculty at Institute of Literature, Shanghai Academy of Social Sciences. Major research interest: computer art and aesthetics.

## I. Introduction

Ever since the invention of radio broadcasting, modern technologies have brought never-seen convenience to the distribution of artworks, followed by the introduction of television, videotapes, and the Internet to artistic contexts. Furthermore, these technologies have also penetrated the process of artistic creation at all stages, converting and renewing materials, workflows, and morphoses of artworks. It is now evident that no insights into today's art and artworks without concerns regarding technical factors, among which computer technology is the foremost. Artistic genres and movements including 'media art (Medienkunst),' 'digital art,' 'generative art,' and 'algorithmic art' are emerging, and a handful of specialized institutions for art collection, exhibition, and commerce have come into existence. The recent decade sees the prosperity of artificial intelligence, which carries out the very genre of 'artificial intelligence art' (AI art). Controversial and heated as these modes of artistic practice are, the discussions relating to them are however ambiguous for the lack of a common concept to denote them in their particularity and commonality.

Tracing back to the late 1950s, several artists and theorists proposed the term of 'computer art,' which is the common ancestor of 'media art' etc. Owing to the literal meaning of the phrase,

'computer art' is plausibly a right term for art practice involving computer technology. However, comprehending the term solely as 'art involving computers' could be meaningless in contemporary artistic situations: movie effects and computer-generated (CG) animations are now commonplace in filming, and utilizing computers for daily business or coordinating artists is no longer an exceptional case. Evidently, 'computer art' would make no distinctions and therefore senseless if we denote all artistic practice involving computers with the term. A more distinguishable, more specific, and more coherent delimitation is required for the renovation of the concept of computer art, which could contribute to commeasurable discussions regarding that newly emerging series in artistic practice.

The present paper serves as an introductory essay to the research of the problematic of computer art, consisting of three major parts. In the first part, a retrospective review of past attempts in defining computer art is presented. It is argued that two formerly prevalent methods, i.e. pragmatic definition and morphological definition, cannot hold for the relentlessly changing practice relating to computer art. The second part proposes an alternative approach – thematic definition, and three themes in computer art are identified. The third part articulates the interrelationships between the triad of themes, and it is demonstrated that the three themes, as primitive problems, form a holistic problematic, and the domain is aesthetic. To avoid confusion, before we reach our definition of computer art, we will use the phrase 'computer-related art' referring to all probable candidates for computer art (works, practices, etc.) and abbreviated as 'CRA' in the text.

**II. Methods of Defining Computer Art**

Some authors confine computer art to the works of those who explicitly mentioned 'computer art' or 'computer graphics' before the 1980s. This methodology can be called 'pragmatic definition' in reference to linguistic pragmatics. Straightforward and clear as it may seem, such definition presumes a questionably coherent usage among those who referred to the term in their discourses. Actually, some of the pioneers in CRA like A. Michael Noll, then an engineer at Bell Laboratories, did agree to the term. However, Noll used the term only in the sense that his works could be seen as paintings, for the reason that his computer-generated pictures could resemble those made by modernist painters.[1] Computer art is in his sense a variant – a hardly qualified imitator of paintings. On the other hand, the 'Stuttgart Group' (Stuttgarter Gruppe) influential in Germany and continental Europe, regarded computer art as a test field for their theoretical ambitions.[2] Exemplified by its founder Max Bense, their comprehension of the term was rooted in the 'informatic aesthetics' (informationstheoretische Ästhetik), which rephrased traditional aesthetic theories with technical terms from computer science (or cybernetics in his time). Therefore, the term 'computer art' means much differently for Stuttgart Group from what Noll had figured out from 'computer art.'

---

[1] A. Michael Noll, "The Beginnings of Computer Art in the United States: A Memoir." *Leonardo*, vol. 27, no. 1, 1994, p. 39.
[2] Cf. Max Bense, *Aesthetische Information: Aesthetica II*. Baden-Baden: Agis-Verlag, 1965.

Meantime, not all practitioners agreed to use the term 'computer art.' Take British painter Harold Cohen for example. Considered an excellent exemplar for computer artists for decades, he was famous for AARON, a painting robot made by him. But computer art to Cohen is 'very reactionary, very old-fashioned.' –

> Because it hasn't changed in any fundamental way in twenty years and that seems to me a sure sign of infertility. Because I've rarely seen any that I didn't find simple-minded and boring. Because so much of it is done by the technology and not by the artist. Because… (when I get started on this particular topic I've been known to go on for hours.) [1]

In Cohen's opinion, he was just 'painting in another way,' and he only reluctantly accepted the term 'computer art' for the sake of public acceptance. Similarly, Frieder Nake, German mathematician and also a well-known pioneer for computer art, proclaimed that 'there should be no computer art' in his 1971 article. According to him, computer-generated artifacts did not change the structure of aesthetic behaviors in comparison with existing artworks and therefore did not form a separate category of art. It is also noteworthy to mention that Cohen and Nake, as opponents against 'computer art,' also diverge in their objectives. While Cohen conceived his works as paintings, Nake urged the audience to 'forget art' and to divert to the new situation where computers could play a vital role in human society. His major interests laid on the changing relationships among creators, computers, compositions, and audiences.

'Computer art,' therefore, would seem ambiguous from the supporters of the term, and self-contradictory and self-denial from the opponents of the term. What complicates the situation is that not all expressions including 'computer art' relate to 'art' in its usual sense. 'Computer Art Contest' held by American trading magazine *Computers and Automation* is known for bringing the term 'computer art' to the general public for the first time in 1964.[2] Nonetheless, the contest was about graphical presentation of state-of-art technologies, and the works were mostly judged from aspects like technical completeness and complexity rather than cultural or artistic significance, at least in its early stage. Similar usage of 'art' existed at that time. Donald E. Knuth, having published his series of studies in computer science titled *The Art of Computer Programming* by the end of the 1960s, stressed that the word 'art' in his title should be comprehended from an 'aesthetic' aspect.[3]

---

[1] Harold Cohen, "Off the Shelf," *The Visual Computer*, vol. 2, no. 3, 1986, pp. 191-94.
[2] Grant D. Taylor, *When the Machine Made Art*, London: Bloomsbury Academic, 2014, p. 27.
[3] Donald E. Knuth, "Computer Programming as an Art." *Communications of the ACM*, vol. 17, no. 12, 1974, pp. 667-673.

Nevertheless, no matter how those technical experts have claimed, programming does not count as 'art.'

Convinced that it is difficult, if not unfeasible, for the pragmatic definition to reconcile the contradictories among different discourses relating to the term 'computer art,' some other scholars proposed alternative definitions by summarizing morphological characteristics of CRA with a selected collection of artworks and expounding analytically from these features. This method can be called 'morphological definition.' For instance, Canadian philosopher Dominic Lopes defines computer art as interactive art running on a computer and 'it's interactive because it's run on a computer.'[1] Interactivity, as a leitmotif in his researches concerning computer art, serves as a starting point in his arguments and gives the systematic nature to his monograph. However, as is shown in Noll's case, computer art in its early stage rarely resort to interactivity.

Prior to Lopes, Nicks Lambert adopted the definition that computer art is a kind of visual art produced with computers in his dissertation,[2] and the definition is thereby compatible with the works by early computer artists. Again, this definition is too broad to measure CRA today, as the style of geometric curves dominant in early CRA-works is common in today's art and design. There is always an 'anachronism' when morphological definitions meet reality.

Notably, such anachronism in the morphological definition is not completely owing to the author's limited horizon when selecting representative CRA-works. Although a number of works, persons, and influential theories have emerged in the relatively short history of all CRA, it is still difficult to extract 'stable' morphological characteristics for the sake of computer art, since at any rate 'computer art' lacks exemplars that have been tested by history and the public. Indeed, computer art is known for its variability in forms, thus making anachronism almost a necessary implication in morphological definition, for its incapability of coping with the historical dimension.

In reply to the above-mentioned situation, some art critics simply call computers 'meta-media' or 'unstable media,' as is figured by Arjen Mulde:

> Unstable media on the other hand are themselves transient, time-based, once-only. They do not deliver a durable structure; they initiate a process. They transfer their consumers to another realm or space or mentality than that of mortality, of ordinary life.[3]

It must be conceded that even 'stable media' like easel paintings encourage innovation, and artists are striving to break conventions for new ways of expression. But in such cases, the alterations or the breakthroughs are made possible only when compared to an established and well-received tradition. A detective story *should* reveal the murderer at the end, a symphony *should* consist of four movements; the conventions can be broken, at the cost that an accompanying justification should be provided in order to 'continue' the tradition with the newly alternated form. In computer art or CRA in general, the situation is quite different. As the unconformity between Lambert and Lopes shows, scholars disagree at the morphological level: they cannot even reach an agreement on the basic form to which computer art should subordinate (Lambert prefers images, while Lopes focuses on interactive installations), not to mention their particular choices of morphological characteristics. Thus, we may conclude that morphological definitions fail (1) to reflect changes in historical development, and (2) to accommodate a variety of heterogeneous practices at the same time.

---

[1] Dominic McIver Lopes, *A Philosophy of Computer Art*. London: Routledge, 2010, p. 27.
[2] Cf. Nicks Lambert, *A Critical Examination of Computer Art*. Doctoral Dissertation, Oxford University, 2003.
[3] Arjen Mulde, "Unstable Media," in *Book for the Electronic Arts*, 2000, available online at https://v2.nl/archive/articles/unstable-media.

The pragmatic definition focuses on the history of the term 'computer art,' especially its early stages; the morphological definition reflects more upon the morphological characteristics of the relevant artistic practice at a certain period of time. Practitioners and researchers have given respectively from their perspectives various definitions with few overlaps, for each of these formerly proposed definitions is reflections of the author's own induction or assertion, which in turn filters the heterogenous modes of practice into homogenous continuity, and the very characteristics of computer art, like variability, are lost. The rapid succession of modes of practice in CRA, along with its intuitively hinted identity, is now calling for a balance between both historical and logical perspectives, so as to portray a more complete picture of computer art.

## III. Thematic Definition and the Three Themes

Similar to the definition of computer art, the later 'media art' encountered blurred boundaries in defining its scope too. In this regard, Christiane Paul, Geert Lovink, Anke Hoffmann, Yvonne Volkart, and many other well-known critics have defined the media art as one made by those 'who develop a personal reflection on the impact of media and digital technologies on our society, rather than with respect to the media used in their practice.'[1]   This definition keeps the term 'media art' sufficiently open to accommodate its still-evolving historical process, while at the same time delineating a relatively clear conceptual boundary. This is called a thematic definition. By doing so, the discussion regarding 'media art' can be quickly and clearly directed, maintaining a considerable degree of commensurability among discourses.[2] In this way, 'media art' gains a stronger impact and recognition, compared to other terms such as 'net art,' 'algorithmic art,' etc.,

When traditional methods of definition, such as pragmatic definition and morphological definition, fail to give satisfactory results, the thematic definition becomes a promising alternative. It can liberate the work from morphological examination and return it to its original historical conditions, together with the linguistic history of the particular discourse that calls this artistic activity and work. In this approach, the form of the work is seen as the result of a prior process of creation, and the socio-historical context, the personal choices of the artist, and the form of the product can all be brought into the scope of the 'subject's' concerns. Indeed, Lev Manovich, a scholar of contemporary media theory, has already adopted this thematic methodology unconsciously in describing computer art as 'research into new aesthetic possibilities of new media.'[1]

---

[1] Boris Magrini, *Confronting the Machine: An Enquiry into the Subversive Drives of Computer-Generated Art*. Berlin: De Gruyter, 2017, p. 250.
[2] Cf. Thomas S. Kuhn, *The structure of scientific revolutions*. 3rd ed. Chicago: University of Chicago Press, 1996, p. 148.

Magrini argues that the thematic definition of 'media art' introduces too vague a sense of 'reflection,' thus giving the researcher too much room for interpretation so that it is almost possible to arbitrarily interpret certain works as 'media art' by including or excluding them from the scope of 'reflection.' However, this does not mean that thematic definitions are necessarily imprecise as a method for defining. Media art theorists have constructed a 'digital divide' to reinforce what they identify as the theme of reflections on media, arguing that 'media art' in the strictest sense of the term is repelled by contemporary art. In fact, major contemporary art exhibitions have included works that have been labeled as media art at least since the 1990s. Flaws in definition often appear in conflict with facts, and the excessive room for interpretation and arbitrariness is ultimately due to a lack of precise grasp of the factual materials and a lack of awareness of one's own research perspective. Popper's situational logic is a good remedy for these lapses.[2]

Situational logic holds that the researcher is always limited by his or her own perspective and must therefore be aware that the explanation he or she gives is only one of many possible, feasible, and even arguable explanations; on the premise of this full consciousness, he or she traverses through the historical material as extensively as possible, reconstructs the problem situation faced by the historical actors; finally, in the reconstructed problematic situation, he or she arrives at an understanding of other's choices.[3] Fortunately, the practitioners of computer art were also aware of the problematic situations they faced from the very beginning of their works and were therefore particularly keen to leave their own records.

Herbert W. Franke, one of the first theorists for computer art, wrote in his 1971 preface to *Computer Graphics – Computer Art*:

> The works from computers nowadays covered by the term computer art are in my opinion

---

[1] Lev Manovich, "The death of computer art," *The Net Net*. 1996-10-22/2020-09-09. http://absoluteone.ljudmila.org/lev_manovich.php
[2] Karl R. Popper, *The Poverty of Historicism*, New York: Harper Collins Publishers, 1964, pp. 149-150.
[3] Cf. Noretta Koertge, "Popper's Metaphysical Research Program for the Human Sciences," *Inquiry*, vol. 18, no. 4, 1975, pp. 437-462.

among the most remarkable products of our time:

- not because they surpass, or even approach, the beauty of traditional forms of art, but because they place established ideas of beauty and art in question;

- not because they are intrinsically satisfactory or even finished, but because their very unfinished form indicates the great potential for future development;

- not because they resolve problems, but because they raise and expose them. [1]

Coincidentally, Ruth Leavitt, curator and a former abstract expressionist painter, enlisted dozens of questions in her invitation for contributions to a handful of computer artists, including not only the invitee's knowledge background and creation motive, but also several questions specifically referring to computer art:

Could your work be done without the aid of a computer? If yes, why use the computer?

To what extent are you involved in the technical production of your work, for example, in programming?

Do you feel art work created with a computer has now or will have an impact on art as a whole in the future?

Do you intend to continue using the computer to create art pieces?

Do you recommend the use of the computer for others in creating works of art? [2]

These two instances may give us an insight into the vitality of theoretical (aesthetic) problems and the awareness of the practitioners. The 'problem-oriented' character in computer art has generated and kept valuable materials for situational logic. Practitioners of computer art have communicated their concerns to audiences through their work, along with a large number of critics and researchers whose concerns have intersected to a great extent. The following three themes emerge from their legacies and historical documents.

Firstly, the relationship between art and technology. In 1969 International Symposium on Computers and Visual Research, the experts declared in their 'Zagreb manifesto' that

It is now evident that, where art meets science and technology, the computer and related disciplines provide a nexus. […] The aesthetic demands of artists necessarily lead them to seek an alliance with the most advanced research in natural and artificial intelligence. […] [P]eople welcome the insight of the artist in this context, lest we lose sight of humanity and

---

[1] Herbert W. Franke, *Computer Graphics – Computer Art*. Schrack Antje, trans. Berlin: Springer, 1985, p. ix.
[2] Ruth Leavitt, *Artist and Computer*, New York: Harmony Books, 1976, p. vii.

beauty.[1]

For the term 'computer art' refers simultaneously to 'computer (technology)' and 'art,' it is perfectly a justifiable candidate for such nexus. From the question whether people are over-reliant on the Internet to the topic on the development of armed drones, contemporary practitioners of computer art articulate their views through their work from their own experiences and observations with customized hardware, embedded devices, and even drones and 3D printers. This evidence imply that the relationship between art and technology is the central concern among all problems.

In 1930s, Paul Valéry predicted the application of novel technologies in art, especially in the distribution of artworks.[2] Following his inspiration, Benjamin, Brecht, among others, further discussed and assumed future art-technology relationship, drawing conclusions to emphasize the importance of universal and bidirectional communication. These predictions only come into existence after the invention of the computer and the Internet. Contemporary scholars including Kittler,[3] Hansen,[4] Manovich,[5] and others, provides insights into the changes that technology has brought with computer software and networks that make it possible, and how it has influenced and reflected the public's understanding of the relationship between art and technology. As Noël Carroll summarizes:

> [I]f we have in mind a narrower conception of technology, one that pertains to the routine, automatic, mass production of multiple instances of the same product-be they cars or shirts- then the question of the relation of art to technology is a pressing one for our century. For in our century, especially, traffic with artworks has become increasingly mediated by technologies in the narrower (mass production/distribution) sense of the term.[6]

And in alliance with another tradition that can be traced back to Max Scheler and Oswald Spengler, many scholars, rooted in the ideology of conservatism, have also derived a social-critical dimension from the discussion of the relationship between art and technology. Heidegger's 'The Origin of the Work of Art' (written in 1935-1937 and published in 1950) is among most representative works. He was not only critical of the development of modern technology, but also keenly aware of the kinship between art and technology. It is worth noting, however, that the term 'technology' has evolved from an all-encompassing, comprehensive concept, like Heidegger's, to a more specific conception of computer technology in contemporary society recently. Computer technology has not only become a prominent representation of today's technical achievement, but has also been equated with 'technology' itself, while artworks are constantly naturalizing technology and providing technology as an iconography.[1] In this 'postmodern' situation, the practice of computer art and the theme of art-technology relationship constitute and shape each other. Through the 'social' variable latent in art as well as in technology, computer art has also become an important way for artists to intervene in reality.

---

[1] Gordon Hyde, Jonathan Benthall, and Gustav Metzger, "Zagreb Manifesto," *Bit International*, vol. 7, 1969, p. 298.
[2] Paul Valéry, "La Conquête de l'ubiquité," in *Œuvres*, vol. II, Paris: Gallimard, 1960, pp. 1283-1287.
[3] Friedrich Kittler, "There Is No Software," in *Literature, Media, Information Systems: Essays*. John Johnston, ed. Amsterdam: Overseas Publishers Association, 1997, pp. 147-155.
[4] Cf. Mark B.N. Hansen, *New Philosophy for New Media*, Cambridge, Massachusetts: MIT Press, 2006.
[5] Cf. Lev Manovich, *Software Takes Command*, New York: London: Bloomsbury Academic, 2013.
[6] Noël Carroll, "The Ontology of Mass Art," *The Journal of Aesthetics and Art Criticism*, vol. 55, no. 2, 1997, pp. 187-199.

Secondly, the machine creation problem, which usually associated with advances in artificial intelligence. In 2018, a 'painting' claimed to be 'created by artificial intelligence' was sold at a mainstream art auction house, causing an immediate uproar. A series of discussions about whether 'artificial intelligence' is, or can be, 'creative,' whether it constitutes a 'creative subject,' and even whether computer-aided creation is ethical, are prevailing on periodicals and newspapers, giving an obvious example for the problem.

The origins of the machine creation problem can be traced back to more than a century before the birth of the modern electronic computer. In 1842, British mathematician Lady Lovelace anchored the machine for 'symbolic (operative) computation' to the creation of art when analyzing and deciphering the mathematical equivalent of modern computers – the sketch of the Babbage's Analytical Engine. She explicitly identifies part of artistic creation, such as musical arranging, as a task that can be done by Analytical Engine, while excluding the possibility of 'originating.'[2] Turing in his 'Computational Machine and Intelligence,' symbolic of the birth of artificial intelligence as an engineering field, responded to Lady Lovelace's assertions, and made extensive discriminations in defending the possibility for 'machine intelligence.'[3] From the 1970s to the present, Margaret A. Boden, a longtime authority on the philosophy of artificial intelligence, has repeatedly addressed the question of whether machines can be original and can be endowed with 'creativity,' arguing that the legitimacy of the creation of art by computers lies in the possibility of the computer modeling human creativity for artistic creation.[4]

The machine creation problem links the theme of computer art with the development of technology. As far back as the 1950s, in the midst of the cybernetic boom, people began to ask whether artistic creation involving computers could be justified as 'art,' constantly denying or supporting the machine's creative ability with concepts like 'creativity' or 'subjectivity.'[5] In the 1970s, when Harold Cohen was constructing his robot, AARON, he also made it clear that the project is fully associated with artificial intelligence.[1] His contemporary artist Edward Ihnatowicz was also an active figure in technical discussions, writing extensively on perceptions and intelligence.[2] For decades, almost every boom in computer technology, and in particular artificial intelligence technology, has revived the debate about whether machines can create art. This fact establishes the importance of the machine creation problem for computer art.

---

[1] Paul Crowther, *Geneses of Postmodern Art: Technology as Iconology*, New York: Routledge, 2019, p. 3.
[2] Ada Countess Lovelace, "Translator's Notes to an Article on Babbage's Analytical Engine," *Scientific Memoirs*, 1842, vol. 3, pp. 691-731.
[3] Alan M. Turing, "Computing Machinery and Intelligence," *Mind*, vol. 59, no. 236, 1950, pp. 433-460.
[4] Margaret Boden, *The Creative Mind*, New York: Routledge, 2003, pp. 16-17; Margaret A. Boden, Computer Models of Creativity, *AI Magazine*, 2009, 30 (3), p. 23.
[5] See, for example, Han Wei (韓偉), "On the aesthetic trouble of today's AI literature (論當下人工智能文學的審

Thirdly, the problem of the ontology of art. Be the machine creation problem bounded with the development of technology, the ontology of art forms the focus of discussion in the field of aesthetics, and it has thus become the third theme of computer art. Early theories of computer art, represented by Bense, attempted for long to reconstruct the provisions of relevant traditional aesthetics concerning the nature of art and works of art by introducing technical terminology, especially informatics. Although this discursive scheme itself has proved to be a failure, it also shows that these practitioners and researchers are aware of the serious impact of computer technology on traditional notions of the work of art and its essence.[3]

The problem of the ontology of art is an issue almost as old as philosophy itself. In the context of contemporary scholarship, it still holds a unique fascination. Not only Heidegger and other Continental scholars revisit the problem in the context of rapid technological changes – Heidegger's analysis of art is always inseparable from the discussion of 'thing' (Ding), and always returns to 'being' (Sein). – Anglo-Saxon analytic aestheticians are also active in discussing the ontology of art. Gregory Currie and Richard Wollheim have developed two opposing theories in response to the growing phenomenon in modern and postmodern artworks. The latter's notion of multiple ontologies argues that different categories of art have different ontological properties, and that the ontology of art is an enumeration of these different categories,[4] while the former's theory of artistic action equates a work of art with an action type, marked by a particular action performed by a particular person on a particular occasion,[5] paving the way to Dave Davies's recent theory that art is performance.[6] Davies extrapolated the theory of action types to its extreme, and argued that images such as paintings are only some kind of evidence and products of the artist's painting process, and that the real work of art is the 'performance' of the artist's painting process. Similarly, a number

---

美困境)", *Wenyi Zhengming* (文藝爭鳴), no. 7, 2020, pp. 100-106.
[1] Grant D. Taylor, *When the Machine Made Art*, London: Bloomsbury Academic, 2014, p. 127.
[2] Cf. Edward Ihnatowicz, "Towards a Thinking Machine," in *Artist and Computer*, New York: Harmony Books, 1976, pp. 32-34; Edward Ihnatowicz, "The Relevance of Manipulation to the Process of Perception," *Mathematics and its Applications*, 1977, vol. 1, pp. 133-135.
[3] Frieder Nake, "Information Aesthetics: An Heroic Experiment," *Journal of Mathematics and the Arts*, vol. 6, nos. 2-3, 2012, pp. 65-75.
[4] Richard Wollheim, *Art and Its Objects*, Cambridge: Cambridge University Press, 1980.
[5] Gregory Currie, *An Ontology of Art*, New York: St. Martin's Press, 1989, pp. 7, 79.
[6] Dave Davies, *Art as Performance*, Hoboken: John Wiley & Sons, 2008, p. x.

of computer artists have argued that the truly 'creative part' of their work, and thus the truly 'artistic' part, lies in the process of programming, rather than in the pictures that they produce through programs.

In addition to those general theories of the ontology of art, Paul Crowther devoted a discussion to that of visual art involving computers in particular. He similarly argues that 'digital visual art might move in the direction of musical performance, with the work being increasingly presented as a vehicle for such interpretation rather than an individual visual artwork per se.'[1] John Lechte's discussion concerning 'genealogy and ontology' of digital images, confining also in the field of visual art, emphasized that images are irreducible to material substances, or mental, cultural, social, or symbolic constructions, only to conclude that the ontological status of digital images is still mysterious.[2] In short, practitioners and scholars alike have noticed the impact of computer art on the ontology of art, and are eagerly awaiting the emergence of effective theoretical arguments, but the responses to this problem are still fragmented and intuitive. Practitioners speak through their own works to reanalyze what the meaning of 'art' is and what kind of existential forms artworks in computer art have, which always attracts and calls researchers to ponder on the ontology of computer art in particular.

## IV. The Triad of Themes and the Aesthetic Problematic

In the midst of changing contexts and varying patterns, we find that the main concerns about computer art are concentrated under three themes that have not changed fundamentally over the decades. With respect to the interrelationship of these three themes, it can be observed that:

(1) they are not isolated from each other. There are often ontological presuppositions or inferences underlying the view of the relationship between technology and art, and contemporary ontological questions about art and the relationship between art and technology are twisted together. The possibility, once conceived by Adorno, of revealing the 'truth content' of something produced by seeing through the flaws in technical operations for reproduction,[3] has almost vanished in the precise reproduction through computer technology, and the foundations of discussions concerning mechanical reproduction (technische Reproduzierbarkeit) since Benjamin has shaken. It seems that the subsequent development of the critical approach of the Frankfurt School is now in harmony with the phenomenological approach represented by Heidegger, which was originally at odds reciprocally, as they both contain a dimension of the judgment of technology in the context of

---

[1] Paul Crowther, "Ontology and Aesthetics of Digital Art," *The Journal of Aesthetics and Art Criticism*, vol. 66, no. 2, 2008, pp. 161-170.
[2] John Lechte, *Genealogy and Ontology of the Western Image and Its Digital Future*, New York: Routledge, 2012, p. 3.
[3] Theodor W. Adorno, *Aesthetic Theory*. Hullot-Kentor Robert, trans. London: Contiuum, 2002, p. 129.

'artistic truth.' Likewise, the concept of 'creativity' most scholars invoke repeatedly in response to the machine creation problem, contains also a definition of the nature of art, so that the 'creative process' could be an alias for artistic activity.

(2) In the case of the three themes listed above, they all have a long, academic history, and at the same time derive from the major concerns of practitioners in their direct encounter with computer technology, and thus fall within the scope of the primitive questions in a dual sense.[1] Primitive as the three themes are, they are interdependent and mutually supportive, and their extensive problem pedigree accommodates most questions about computer art – if not all the questions in the past – historical development and logical deduction in relation to CRA.

(3) Therefore, the problem of the ontology of art occupy a basic position, organizing the three themes of computer art into an aesthetic problematic. The interrelationship between the three themes of computer art shows that they constitute a whole, thus demarcating a problematic, that is, the logical space of questions. And only by clarifying what is meant by 'art' and 'art creation' at first, can the art-technology problem and the machine creation problem be clearly stated.

At this point, we have identified three themes in the practice and theoretical discourse related to computer art under various definitions, which, as a holistic problematic, identify a distinguishable class of problematic situations in which computer art can emerge, thus giving it an identity. Although not all of the three themes belong to aesthetics, the practice of computer art/CRA, or strictly speaking, the efforts to make 'art' with computers, is impossible without a bite on the problem of the ontology of art. In fact, even defining computer art, as this paper presents, is based on assumptions at the ontological level of art that are still waiting for clarification – and the very first of which is that to what extent 'computer art' constitutes a category of art. We may, for the moment, only briefly respond to this question.

E.H. Gombrich observed that people's anticipation in art actually constitutes an important dimension in the forming of art genres and categories.[2] Here we call it artistic anticipation. In the first place, artistic anticipation is not a rigid, logical constraint, nor is it inevitable. It functions to indicate the range of highly possible forms or candidates and that can be defended with a considerable cumulation of facts. Secondly, like what is often referred to as anticipation, it is future-oriented, and only imaginable in the unfolding of history. The temporality makes the unfinished state of computer art acceptable as an art category, for it contains still the possibility of its 'coming of age.' Last but not least, psychological as the term may seem, anticipation is by no means a purely mental state. It has an external dimension, so that one can observe and confirm it 'from the outside' in its presentation and, in particular, in the efforts devoted to realizing it.

---

[1] Cf. Zheng Yuanzhe (鄭元者), "Primitive Questions, Academic Faith and the Reconstruction of the Ecological Environment of Aesthetics (原始問題、學術忠誠與美學生態重建)," *Theoretical Studies on Literature and Art* (文藝理論研究), no. 6, 2003, pp. 81-85.
[2] Cf. Ernst Hans Gombrich and Didier Eribon, *Looking for Answers: Conversations on Art and Science*. New York: Harry N. Abrams, 1993, p. 163.

The mere anticipation of computer art does not make it appear immediately, but with the benefit of hindsight, we can listen to the words and observe the deeds and recognize it from historical materials, and find that the causal relationship between practice and theory and this artistic anticipation. Coincidentally, Lady Lovelace, in considering computing machine in relation to art, spoke specifically of the 'power of anticipating analytic relationships or truths,' which is an essential difference between humans and machines. Regardless of whether or not future machines may acquire this ability, the above assertion highlights the importance of 'anticipating' as well. The conscious contemplation of the artistic situation under the conditions of high technology constitutes the reason for actions of the computer art practitioners, and even those who took a 'cancellationist' stance towards 'computer art' respond just in an extreme way to the aesthetic problematic, and thus fulfills the artistic anticipation imposed by computer art as defined above. This is why we can use the term 'computer art' (even if used critically) without falling into nihilism.

Ever since Paul Valéry, quite a few theorists, especially those with an affinity with the artistic reality, believed that technology would bring about a new and separate form of art. This anticipation has become more widespread and echoed with the rise of computer technology, and the creation of computer art is also inseparable from the social situation that has given birth to such anticipation – the anticipation that technology and machines will transform human art and subvert the old aesthetics. The triad of themes represents the three faces of this anticipation – social, technical, and aesthetic.

The problematic – the triad – the artistic anticipation. We don't need to worry about the 'modern' imprint that Heidegger is afraid of – 'the art as expression of human life' (die Kunst als Ausdruck des Lebens des Menschen);[1] contrarily, it serves as a favorable premise to prove that computer art belongs to 'art.' Of course, this is not to imply that returning to the problem situation and grasping its original state is the only reasonable way to understand art. It's just that for computer art, a field whose artistic identity is not yet clear, Heidegger's so-called 'anthropological' or 'anthropocentric', 'humanitarian' viewpoint is precisely a reliable approach with which we can seek truth sincerely from historical facts. It is arguable that computer art, a product of (post-)modernity, infiltrates the multitude of expressive relationships between humanity and contemporary high-tech societies. This helps justify computer art as a category of art.

---

[1] Martin Heidegger, *Holzwege*, Hrsg. F.-W. von Herrmann. Frankfurt a.M.: Klostermann, 1950, S. 75.

## V. Conclusion

Thematic definition with situational logic reduces the consensus (on which the pragmatic definition depends) to the historical process of statement and refutation, and the formal characteristics (on which the morphological definition depends) to the art of shaping and sublating forms in practice, so that the emergence and disappearance of consensus and forms are on the chain of 'becoming' where the themes get developed. Therefore, different from the pragmatic definition and the morphological definition which mainly rely on logical speculation, the thematic definition pinpoints the art in the actual historical development process of the problem situations. The true 'power of art' comes from people's concerns about the problem situation they face, and the anticipations that people place on it are evidence of the relevance of this situation. The three themes of computer art are combined to construct an aesthetic problematic with rich connotations. This problematic can endure for a long time. On the one hand, it is due to its broad academic prospects and close relation to modern life. On the other hand, it is precisely because of the long-lasting anticipations.

People have anticipations of computer art – be they the practitioners who personally participate in it, or observers who look at it at distance, or the general public who hear about it. Anticipations entail questions, and questions fall into the aesthetic problematic. The ever-renewing anticipations maintain the drive to explore the issues, promoting the creation, spread, and interpretation of computer art. The existence of computer art as a category of art obtains its premilitary evidence from this fact.